\documentclass[11pt]{revtex4}    
\usepackage{amssymb,epsf}
\usepackage{latexsym}
\begin{document}

\title{Magnetic String Coupled to Nonlinear Electromagnetic Field}
\author{S. H. Hendi\footnote{hendi@mail.yu.ac.ir}}

\address{Physics Department,
College of Sciences, Yasouj University, Yasouj
75914, Iran\\
Research Institute for Astrophysics and Astronomy of Maragha
(RIAAM), P.O. Box 55134-441, Maragha, Iran}

\begin{abstract}
We introduce a class of rotating magnetically charged string
solutions of the Einstein gravity with a nonlinear electrodynamics
source in four dimensions. The present solutions has no curvature
singularity and no horizons but has a conic singularity and yields
a spacetime with a longitudinal magnetic field. Also, we
investigate the effects of nonlinearity on the properties of the
solutions and find that for the special range of the nonlinear
parameter, the solutions are not asymptotic AdS. We show that when
the rotation parameter is non zero, the spinning string has a net
electric charge that is proportional to the magnitude of the
rotation parameter. Finally, we use the counterterm method
inspired by AdS/CFT correspondence and calculate the conserved
quantities of the solutions.
\end{abstract}

\pacs{04.20.Jb, 04.40.Nr, 04.50.+h}
\maketitle

\address{Physics Department,
College of Sciences, Yasouj University, Yasouj
75914, Iran\\
Research Institute for Astrophysics and Astronomy of Maragha
(RIAAM), P.O. Box 55134-441, Maragha, Iran}

\section{Introduction\label{Intro}}

Topological defects are inevitably formed during phase transitions
in the early universe, and their subsequent evolution and
observational signatures must therefore be understood. The string
model of structure formation may help to resolve one of
cosmological mystery, the origin of cosmic magnetic fields
\cite{VachVil91}. There is strong evidence from all numerical
simulations for the scaling behavior of the long string network
during the radiation-dominated era. Apart from their possible
astrophysical roles, topological defects are fascinating objects
in their own right. Their properties, which are very different
from those of more familiar system, can give rise to a rich
variety of unusual mathematical and physical phenomena
\cite{VilenkinBook}.

On another front, nonlinear electromagnetic fields are subjects of
interest for a long time. For example, there has been a renewed
interest in Born-Infeld gravity ever since new solutions have been
found in the low energy limit of string theory. Static and
rotating solutions of Born-Infeld gravity have been considered in
Refs. \cite{Demi86,DehSed,DAH08DH07}.

In this paper, we turn to the investigation of spacetimes
generated by static and spinning string sources in
four-dimensional Einstein theory in the presence of a nonlinear
electromagnetic field which are horinzonless and have nontrivial
external solutions. The basic motivation for studying these kinds
of solutions is that they may be interpreted as cosmic strings.
Cosmic strings are topological structure that arise from the
possible phase transitions to which the universe might have been
subjected to and may play an important role in the formation of
primordial structures. A short review of papers treating this
subject follows. Solutions of Einstein's equations with conical
singularities describing straight strings can easily be
constructed \cite{AryForVil86}. One needs only a spacetime with a
symmetry axis. If one then cuts out a wedge then a space with a
string lying along the axis is obtained. A non axisymmetric
solutions of the combined Einstein and Maxwell equations with a
string has been found by Linet \cite{Linet87}. The
four-dimensional horizonless solutions of Einstein gravity have
been explored in \cite{Vil,Ban}. These horizonless solutions
(\cite{Vil,Ban}) have a conical geometry; they are everywhere flat
except at the location of the line source. The spacetime can be
obtained from the flat spacetime by cutting out a wedge and
identifying its edges. The wedge has an opening angle which turns
to be proportional to the source mass. The extension to include
the Maxwell field has also been done \cite{Bon}. Static and
spinning magnetic sources in three and four-dimensional
Einstein-Maxwell gravity with negative cosmological constant have
been explored in \cite{Lem1,Lem2}. The generalization of these
asymptotically AdS magnetic rotating solutions to higher
dimensions has also been done \cite{Deh2}. In the context of
electromagnetic cosmic string, it has been shown that there are
cosmic strings, known as superconducting cosmic strings, that
behave as superconductors and have interesting interactions with
astrophysical magnetic fields \cite{Wit2}. The properties of these
superconducting cosmic strings have been investigated in
\cite{Moss}. Solutions with longitudinal
and angular magnetic field were considered in Refs. \cite%
{Dehhorizonless1,Dehhorizonless2,DB07,DHWorm}. Similar static
solutions in
the context of cosmic string theory were found in Ref. \cite{VilBanBanSen96}%
. All of these solutions \cite%
{Dehhorizonless1,Dehhorizonless2,DB07,VilBanBanSen96,LeviCivita19Marder58}
are horizonless and have a conical geometry; they are everywhere
flat except at the location of the line source. The extension to
include the electromagnetic field has also been done
\cite{Mukh38Witt85,DiasLemos02}. The generalization of these
solutions in Einstein gravity in the presence of
a dilaton and Born-Infeld electromagnetic fields has been done in Ref. \cite%
{HendiJMP08DSH08}.

Another example of the nonlinear electromagnetic field is
conformally invariant Maxwell field. In many papers,
straightforward generalization of the Maxwell field to higher
dimensions one essential property of the electromagnetic field is
lost, namely, conformal invariance. Indeed, in the
Reissner-Nordstr\"{o}m solution, the source is given by the
Maxwell action which enjoys the conformal invariance in four
dimensions. Massless spin-$1/2$ fields have vanishing classical
stress tensor trace in any dimension, while
scalars can be \textquotedblleft improved\textquotedblright\ to achieve $%
T_{\alpha }^{\alpha }=0$, thereby guaranteeing invariance under
the special
conformal (or full Weyl) group, in accord with their scale-independence \cite%
{DesSch}. Maxwell theory can be studied in a gauge which is
invariant under conformal rescalings of the metric, and at first,
have been proposed by Eastwood and Singer \cite{EastSing}. Also,
Poplawski \cite{Popl08} have been showed the equivalence between
the Ferraris--Kijowski and Maxwell Lagrangian results from the
invariance of the latter under conformal transformations of the
metric tensor. Quantized Maxwell theory in a conformally invariant
gauge have been investigated by Esposito \cite{Espo97}. In recent
years, gravity in the presence of nonlinear and conformally
invariant Maxwell source have been studied in many papers
\cite{HessMart07MHMar,HendiRast}.

The outline of our paper is as follows. We give a brief review of
the field equations of Einstein gravity in the presence of
cosmological constant and nonlinear electromagnetic field in Sec.
\ref{Basic}. In Sec. \ref{Static} we present static horizonless
solutions which produce longitudinal magnetic field, compare it
with the solutions of the standard electromagnetic field and then
investigate the properties of the solutions and the effects of
nonlinearity of the electromagnetic field on the deficit angle of
the spacetime. Section \ref{Angul} will be devoted to the
generalization of these solutions to the case of rotating
solutions and use of the counterterm method to compute the
conserved quantities of them. We finish our paper with some
concluding remarks.

\section{Basic Field Equations\label{Basic}}

Our starting point is the four-dimensional Einstein-nonlinear
Maxwell action
\begin{eqnarray}
I_{G} &=&-\frac{1}{16\pi }\int_{\mathcal{M}}d^{4}x\sqrt{-g}\left(
R-2\Lambda
-\alpha F^{s}\right) -\frac{1}{8\pi }\int_{\partial \mathcal{M}}d^{3}x\sqrt{%
-\gamma }\Theta (\gamma ),  \label{Act}
\end{eqnarray}%
where ${R}$ is the scalar curvature, $\Lambda $ is the
cosmological constant, $F$ is the Maxwell invariant which is equal
to$\ F_{\mu \nu }F^{\mu \nu }$(where $F_{\mu \nu }=\partial _{\mu
}A_{\nu }-\partial _{\nu }A_{\mu }$ is the electromagnetic tensor
field and $A_{\mu }$ is the vector potential), $\alpha $ and $s$\
is a coupling and arbitrary constant respectively. The last term
in Eq. (\ref{Act}) is the Gibbons-Hawking surface term. It is
required for the variational principle to be well-defined. The
factor $\Theta $ represents the trace of the extrinsic curvature
for the boundary ${\partial \mathcal{M}}$ and $\gamma $ is the
induced metric on the boundary. Varying the action (\ref{Act})
with respect to the gravitational field $g_{\mu \nu }$ and the
gauge field $A_{\mu }$, yields
\begin{equation}
G_{\mu \nu }+\Lambda g_{\mu \nu }=T_{\mu \nu },  \label{FE1}
\end{equation}
\begin{equation}
\partial _{\mu }\left( \sqrt{-g}F^{\mu \nu }F^{s-1}\right) =0.  \label{FE2}
\end{equation}%
In the presence of nonlinear electrodynamics field, the
energy-momentum tensor of Eq. (\ref{FE1}) is
\begin{equation}
T_{\mu \nu }=2\alpha \left[ sF_{\mu \rho }F_{\nu }^{\rho }F^{s-1}-\frac{1}{4}%
g_{\mu \nu }F^{s}\right] .  \label{TT}
\end{equation}%
The conserved mass and angular momentum of the solutions of the
above field equations can be calculated through the use of the
substraction method of Brown and York \cite{BY}. Such a procedure
causes the resulting physical quantities to depend on the choice
of reference background. A well-known method of dealing with this
divergence for asymptotically AdS solutions of Einstein gravity is
through the use of counterterm method inspired by AdS/CFT
correspondence \cite{Mal}. In this Letter, we deal with the
spacetimes with zero curvature boundary, $R_{abcd}(\gamma )=0$,
and therefore the counterterm for the stress energy tensor should
be proportional to $\gamma ^{ab}$. We find the suitable
counterterm which removes the divergences as
\begin{equation}
I_{ct}=-\frac{1}{4\pi }\int_{\partial \mathcal{M}}d^{3}x\frac{\sqrt{-\gamma }%
}{l}.  \label{cont}
\end{equation}%
Having the total finite action $I=I_{G}+I_{\mathrm{ct}}$, one can
use the quasilocal definition to construct a divergence free
stress-energy tensor \cite{BY}. Thus the finite stress-energy
tensor in four-dimensional Einstein-nonlinear Maxwell gravity with
negative cosmological constant can be written as
\begin{equation}
T^{ab}=\frac{1}{8\pi }\left[ \Theta ^{ab}-\Theta \gamma
^{ab}+\frac{2\gamma ^{ab}}{l}\right] .  \label{Stres}
\end{equation}%
The first two terms in Eq. (\ref{Stres}) are the variation of the action (%
\ref{Act}) with respect to $\gamma _{ab}$, and the last term is
the variation of the boundary counterterm (\ref{cont}) with
respect to $\gamma _{ab}$. To compute the conserved charges of the
spacetime, one should choose
a spacelike surface $\mathcal{B}$ in $\partial \mathcal{M}$ with metric $%
\sigma _{ij}$, and write the boundary metric in ADM
(Arnowitt-Deser-Misner) form:
\[
\gamma _{ab}dx^{a}dx^{a}=-N^{2}dt^{2}+\sigma _{ij}\left( d\varphi
^{i}+V^{i}dt\right) \left( d\varphi ^{j}+V^{j}dt\right) ,
\]%
where the coordinates $\varphi ^{i}$ are the angular variables
parameterizing the hypersurface of constant $r$ around the origin,
and $N$ and $V^{i}$ are the lapse and shift functions,
respectively. When there is a Killing vector field $\mathcal{\xi
}$ on the boundary, then the quasilocal conserved quantities
associated with the stress tensors of Eq. (\ref{Stres}) can be
written as
\begin{equation}
Q(\mathcal{\xi )}=\int_{\mathcal{B}}d^{2}x\sqrt{\sigma }T_{ab}n^{a}\mathcal{%
\xi }^{b},  \label{charge}
\end{equation}%
where $\sigma $ is the determinant of the metric $\sigma _{ij}$, $\mathcal{%
\xi }$ and $n^{a}$ are, respectively, the Killing vector field and
the unit
normal vector on the boundary $\mathcal{B}$. For boundaries with timelike ($%
\xi =\partial /\partial t$) and rotational ($\varsigma =\partial
/\partial \phi $) Killing vector fields, one obtains the
quasilocal mass and angular momentum
\begin{eqnarray}
M &=&\int_{\mathcal{B}}d^{2}x\sqrt{\sigma }T_{ab}n^{a}\xi ^{b},
\label{Mastot} \\
J &=&\int_{\mathcal{B}}d^{2}x\sqrt{\sigma }T_{ab}n^{a}\varsigma
^{b}. \label{Angtot}
\end{eqnarray}%
These quantities are, respectively, the conserved mass and angular
momentum of the system enclosed by the boundary $\mathcal{B}$.
Note that they will both depend on the location of the boundary
$\mathcal{B}$ in the spacetime,
although each is independent of the particular choice of foliation $\mathcal{%
B}$ within the surface $\partial \mathcal{M}$.

\section{Static nonlinear magnetic string \label{Static}}

Here we want to obtain the four dimensional solutions of Eqs. (\ref{FE1})-(%
\ref{TT}) which produce a longitudinal magnetic fields along the
$z$ direction. We assume the following form for the metric
\cite{Lem1}
\begin{equation}
ds^{2}=-\frac{\rho ^{2}}{l^{2}}dt^{2}+\frac{d\rho ^{2}}{f(\rho )}%
+l^{2}f(\rho )d\varphi ^{2}+\frac{\rho ^{2}}{l^{2}}dz^{2}.
\label{Met1}
\end{equation}%
The function $f(\rho )$ should be determined and $l$ has the
dimension of length which is related to the cosmological constant
$\Lambda $ by the relation $l^{2}=-3/\Lambda $. The coordinate $z$
has the dimension of length and ranges $-\infty <z<\infty $, while
the angular coordinate $\phi $ is dimensionless as usual and
ranges in $0\leq \phi <2\pi $. The motivation for
this curious choice for the metric gauge $[g_{tt}\varpropto -\rho ^{2}$ and $%
(g_{\rho \rho })^{-1}\varpropto g_{\phi \phi }]$ instead of the
usual Schwarzschild gauge $[(g_{\rho \rho })^{-1}\varpropto
g_{tt}$ and $g_{\phi \phi }\varpropto \rho ^{2}]$ comes from the
fact that we are looking for
magnetic solutions. Taking the trace of the gravitational field equation (%
\ref{FE1}), the scalar curvature is expressed in terms of the
Maxwell invariant $F$ \ and cosmological constant $\Lambda $\ as
\[
R=2\left[ \Lambda -\alpha (s-1)F^{s}\right] .
\]%
Before studying in details the field equations, we first specify
the sign of the coupling constant $\alpha $ in term of the
exponent $s$ in order to ensure a physical interpretation of our
future solutions. In fact, the sign of the coupling constant
$\alpha $ in the action (\ref{Act}) can be chosen such that the
energy density, i.e. the $T_{_{\widehat{t}\widehat{t}}}$ component
of the energy-momentum tensor in the orthonormal frame, is
positive
\[
T_{_{\widehat{t}\widehat{t}}}=\frac{\alpha }{2}F^{s}>0.
\]%
As a direct consequence, one can show that the Maxwell invariant $F=\frac{2}{%
l^{2}}(F_{\phi \rho })^{2}$ is positive and hence, the sign of the
coupling constant $\alpha $ should be positive, which can be set
to $1$ without loss of generality. It is well-known that the
electric field is associated with the time component, $A_{t}$, of
the vector potential while the magnetic field is associated with
the angular component $A_{\phi }$. From the above facts, one can
expect that a magnetic solutions can be written in a metric gauge
in which the components $g_{tt}$ and $g_{\phi \phi }$ interchange
their roles relatively to those present in the Schwarzschild gauge
used to describe electric solutions. The Maxwell equation
(\ref{FE2}) can be integrated immediately to give
\begin{equation}
F_{\phi \rho }=\left\{
\begin{array}{cc}
0, & s=0,\frac{1}{2} \\
\frac{-2ql^{2}}{\rho }, & s=\frac{3}{2} \\
\frac{2(2s-3)ql^{2}}{(2s-1)\rho ^{2/(2s-1)}}, & \text{otherwise}%
\end{array}%
\right. ,  \label{Ftr}
\end{equation}%
where $q$, an integration constant where the electric charge of
the string is related to this constant for spinning string.
Inserting the Maxwell
fields (\ref{Ftr}) and the metric (\ref{Met1}) in the field equation (\ref%
{FE1}), we can simplify these equations as

\begin{equation}
\rho f^{^{\prime }}(\rho )+f(\rho )+\Lambda \rho ^{2}-H(\rho )=0
\label{Eqrr}
\end{equation}
where%
\begin{equation}
H(\rho )=\left\{
\begin{array}{cc}
0, & s=0,\frac{1}{2} \\
\frac{16q^{3}l^{3}\sqrt{2}}{\rho }, & s=\frac{3}{2} \\
2^{s}(2s-1)\rho ^{2}\left[ \frac{2\left( 2s-3\right) ql}{\left(
2s-1\right)
\rho ^{(s-2)/(2s-1)}}\right] ^{2s}, & \text{otherwise}%
\end{array}%
\right.   \label{Eqrr2}
\end{equation}%
where the \textquotedblleft prime\textquotedblright\ denotes
differentiation with respect to $\rho $. One can show that these
equations have the following solutions
\begin{equation}
f(\rho )=-{\frac{\Lambda \,\rho ^{2}}{3}}+{\frac{2ml^{3}}{\rho
}+}\left\{
\begin{array}{cc}
0, & s=0,\frac{1}{2} \\
\frac{16q^{3}l^{3}\sqrt{2}\left( 1+\ln \rho \right) }{\rho }, &
s=\frac{3}{2}
\\
\frac{2^{3s}\left( 2s-3\right) ^{2s-1}(ql)^{2s}}{2\left(
2s-1\right)
^{2s-2}\rho ^{2/(2s-1)}}, & \text{Otherwise}%
\end{array}%
\right. ,  \label{f(r)}
\end{equation}%
where $m$ is the integration constant which is related to mass
parameter. In the linear case ($s=1$), the solutions reduce to the
asymptotically AdS horizonless magnetic string solutions for
$\Lambda =-3/l^{2}$ \cite{Lem2}. Here, we want to investigate the
effects of the nonlinearity on the
asymptotic behavior of the solutions. It is worthwhile to mention that for $%
0<s<\frac{1}{2}$, the asymptotic dominant term of Eq. (\ref{f(r)})
is third term and the solutions of the Einstein-nonlinear Maxwell
field equations are not asymptotically AdS, but for the cases
$s<0$\ or $s>\frac{1}{2}$ (include of $s=\frac{3}{2}$), the
asymptotic behavior of Einstein-nonlinear Maxwell
field solutions are the same as linear AdS case. Equations (\ref{Ftr})-(\ref%
{f(r)}) show that the magnetic field is zero for the cases $s=0,\frac{1}{2}$%
, and the metric function (\ref{f(r)}) does not possess a charge
term and it corresponds to uncharged asymptotically AdS one.

Now, we want to investigate the special case, such that the
electromagnetic field equation be invariant under conformal
transformation ($g_{\mu \nu }\longrightarrow \Omega ^{2}g_{\mu \nu
}$ and $A_{\mu }\longrightarrow A_{\mu }$). Consider the
Lagrangian of the form $L(F)$, where $F=F_{\mu \nu
}F^{\mu \nu }$. It is easy to show that for this form of Lagrangian in $4$%
-dimensions, $T_{\mu }^{\mu }\propto \left[ F\frac{dL}{dF}-L\right] $; so $%
T_{\mu }^{\mu }=0$ implies $L(F)=Constant\times F$. It is
worthwhile to mention that only for linear case $s=1$, the
electromagnetic field equation is invariant under conformal
transformation.

Here, we want to study the general structure of the solutions. One
can find that the Kretschmann scalar, $R_{\mu \nu \lambda \kappa
}R^{\mu \nu \lambda \kappa }$, is
\[
R_{\mu \nu \lambda \kappa }R^{\mu \nu \lambda \kappa }=\left( \frac{%
d^{2}f(\rho )}{d{\rho }^{2}}\right) ^{2}+4\left( \frac{1}{\rho }\frac{%
df(\rho )}{d{\rho }}\right) ^{2}+4\left( \frac{f(\rho )}{{\rho
}^{2}}\right) ^{2}.
\]%
It is easy to show that the Kretschmann scalar $R_{\mu \nu \lambda
\kappa }R^{\mu \nu \lambda \kappa }$ diverges at $\rho =0$ and
therefore one might think that there is a curvature singularity
located at $\rho =0$. However, as will be seen below, the
spacetime will never achieve $\rho =0$. Now, we look for the
existence of horizons and, in particular, we look for the possible
presence of magnetically charged black hole solutions. The
horizons, if any exist, are given by the zeros of the function
$f(\rho )=(g_{\rho \rho })^{-1}$. Let us denote the largest
positive root of $f(\rho )=0$ by $r_{0}$. The function $f(\rho )$
is negative for $\rho <r_{0}$, and therefore one may think that
the hypersurface of constant time and $\rho =r_{0}$ is the
horizon. However, the above analysis is wrong. Indeed, we
first notice that $g_{\rho \rho }$ and $g_{\phi \phi }$ are related by $%
f(\rho )=g_{\rho \rho }^{-1}=l^{-2}g_{\phi \phi }$, and therefore when $%
g_{\rho \rho }$ becomes negative (which occurs for $\rho <r_{0}$) so does $%
g_{\phi \phi }$. This leads to an apparent change of signature of
the metric from $+2$ to $-2$. This indicates that we are using an
incorrect extension. To get rid of this incorrect extension, we
introduce the new radial coordinate $r$ as
\begin{equation}
r^{2}=\rho ^{2}-r_{0}^{2}\Rightarrow d\rho ^{2}=\frac{r^{2}}{r^{2}+r_{0}^{2}}%
dr^{2}.  \label{Tr1}
\end{equation}%
With this coordinate change the metric (\ref{Met1}) is written as
\begin{equation}
ds^{2}=-\frac{r^{2}+r_{0}^{2}}{l^{2}}dt^{2}+l^{2}f(r)d\phi ^{2}+\frac{r^{2}}{%
(r^{2}+r_{0}^{2})f(r)}dr^{2}+\frac{r^{2}+r_{0}^{2}}{l^{2}}dz^{2},
\label{Met2}
\end{equation}%
where the coordinates $r$ assumes the values $0\leq r<\infty $,
and $f(r)$, is now given as

\begin{equation}
f(r)=-{\frac{\Lambda \,(r^{2}+r_{0}^{2})}{3}+{\frac{2ml^{3}}{%
(r^{2}+r_{0}^{2})^{1/2}}}+}\left\{
\begin{array}{cc}
0, & s=0,\frac{1}{2} \\
\frac{8q^{3}l^{3}\sqrt{2}\left[ 2+\ln \left(
r^{2}+r_{0}^{2}\right) \right]
}{\left( r^{2}+r_{0}^{2}\right) ^{1/2}}, & s=\frac{3}{2} \\
\frac{2^{3s}\left( 2s-3\right) ^{2s-1}(ql)^{2s}}{2\left(
2s-1\right)
^{2s-2}(r^{2}+r_{0}^{2})^{1/(2s-1)}}, & \text{Otherwise}%
\end{array}%
\right. ,  \label{f2}
\end{equation}
The electromagnetic field equation in the new coordinate is
\begin{equation}
F_{\phi r}=\left\{
\begin{array}{cc}
0, & s=0,\frac{1}{2} \\
\frac{-2ql^{2}}{\left( r^{2}+r_{0}^{2}\right) ^{1/2}}, & s=\frac{3}{2} \\
\frac{2(2s-3)ql^{2}}{(2s-1)(r^{2}+r_{0}^{2})^{1/(2s-1)}}, & \text{otherwise}%
\end{array}%
\right. .  \label{f33}
\end{equation}%
One can show that all curvature invariants (such as Kretschmann
scalar,
Ricci scalar, Ricci square, Weyl square and so on) are functions of $%
f^{\prime \prime }$, $f^{\prime }/r$ and $f/r^{2}$. Since these
terms do not diverge in \ the range $0\leq r<\infty $, one finds
that all curvature invariants are finite. Therefore this spacetime
has no curvature singularities and no horizons. It is worthwhile
to mention that the magnetic solutions obtained here have distinct
properties relative to the electric solutions obtained in
\cite{HendiRast}. One can expect magnetic solutions from electric
solution by a double Wick rotation such as $t\longrightarrow i\phi
$ and $\phi \longrightarrow it/l$, ($i=\sqrt{-1}$). Indeed, the
electric solutions have black holes, while the magnetic do not.
However, the spacetime (\ref{Met2}) has a conic geometry and has a
conical singularity at $r=0$, since:
\begin{equation}
\lim_{r\rightarrow 0}\frac{1}{r}\sqrt{\frac{g_{\phi \phi
}}{g_{rr}}}\neq 1. \label{limit}
\end{equation}%
That is, as the radius $r$ tends to zero, the limit of the ratio
\textquotedblleft \emph{circumference/radius}\textquotedblright\ is not $%
2\pi $ and therefore the spacetime has a conical singularity at
$r=0$. The canonical singularity can be removed if one identifies
the coordinate $\phi $ with the period
\begin{equation}
\text{Period}_{\phi }=2\pi \left( \lim_{r\rightarrow 0}\frac{1}{r}\sqrt{%
\frac{g_{\phi \phi }}{g_{rr}}}\right) ^{-1}=2\pi (1-4\mu ),
\label{period}
\end{equation}%
where $\mu $ is given by
\begin{eqnarray}
\mu  &=&\frac{1}{4}\left( 1-\frac{2}{lr_{0}\left( \Omega -2\Lambda \right) }%
\right) ,  \label{miu} \\
\Omega  &=&\left\{
\begin{array}{cc}
0, & s=0,\frac{1}{2} \\
\frac{2^{11/2}q^{3}l^{3}}{r_{0}^{3}}, & s=\frac{3}{2} \\
\frac{8^{s}(2s-1)q^{2s}l^{2s}}{\left( \frac{2s-1}{2s-3}\right)
^{2s}r_{0}^{4s/(2s-1)}}, & \text{otherwise}%
\end{array}%
\right.
\end{eqnarray}%
The above analysis shows that near the origin $r=0$, the metric
(\ref{Met2}) describes a spacetime which is locally flat but has a
conical singularity at $r=0$ with a deficit angle $\delta \phi
=8\pi \mu $. Since near the origin the metric (\ref{Met2}) is
identical to the spacetime generated by a cosmic
string, by using the Vilenkin procedure, one can show that $\mu $ of Eq. (%
\ref{miu}) can be interpreted as the mass per unit length of the string \cite%
{Vil2}.

\begin{figure}[tbp]
\epsfxsize=7cm \centerline{\epsffile{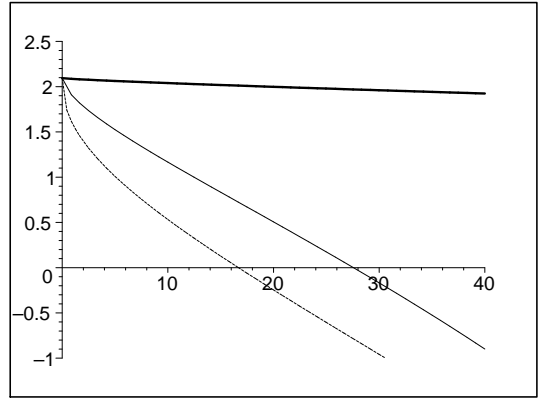}} \caption{The
deficit angle versus $q$ for $r_{0}=0.5$, $l=1$, and $s=0.2 $
(dotted line), $s=0.3 $ (continuous line) and $s=0.4$ (bold
line).} \label{Fig1}
\end{figure}
\begin{figure}[tbp]
\epsfxsize=7cm \centerline{\epsffile{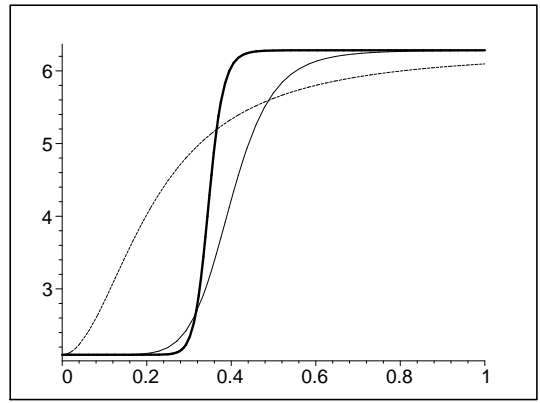}} \caption{The
deficit angle versus $q$ for $r_{0}=0.5$, $l=1$, and $s=1 $
(dotted line), $s=4 $ (continuous line) and $s=10$ (bold line).}
\label{Fig2}
\end{figure}
\begin{figure}[tbp]
\epsfxsize=7cm \centerline{\epsffile{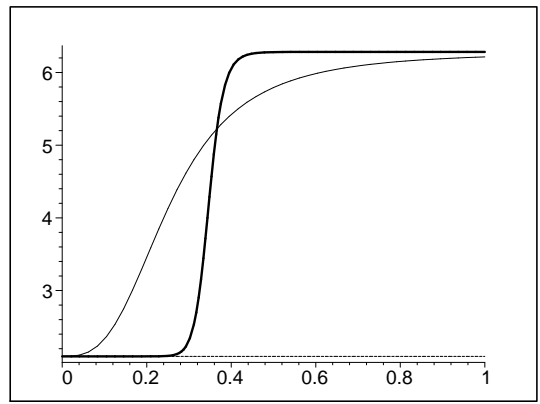}} \caption{The
deficit angle versus $q$ for $r_{0}=0.5$, $l=1$, and $s=0$ or $1/2
$ (dotted line), $s=3/2 $ (continuous line) and $s=10$ (bold
line).} \label{Fig3}
\end{figure}
Also, In order to investigate the effects of the nonlinearity of
the magnetic field on deficit angle $\delta \phi $, we plot it
versus the charge
parameter $q$ in three figures. Fig. \ref{Fig1} shows that for $0<s<%
\frac{1}{2}$, deficit angle decreases as the charge parameter of
the spacetime, $q$, increases. But for a constant value of $q$, as
the nonlinear
parameter, $s$, increases, deficit angle increases too. Also, figures \ref%
{Fig2} and \ref{Fig3} show that for $s>\frac{1}{2}$, deficit angle
increases as the charge parameter of the spacetime, $q$, increases
and as the nonlinear parameter, $s$, increases, the rate of
deficit angle growth increases too.

\section{Spinning nonlinear magnetic string\label{Angul}}

Now, we would like to endow the spacetime solutions (\ref{Met1})
with a rotation. In order to add angular momentum to the
spacetime, we perform the following rotation boost in the $t-\phi
$ plane
\begin{equation}
t\mapsto \Xi t-a\phi ,\hspace{0.5cm}\phi \mapsto \Xi \phi
-\frac{a}{l^{2}}t, \label{Tr}
\end{equation}%
where $a$ is a rotation parameter and $\Xi =\sqrt{1+a^{2}/l^{2}}$.
Substituting Eq. (\ref{Tr}) into Eq. (\ref{Met2}) we obtain
\begin{eqnarray}
ds^{2} &=&-\frac{r^{2}+r_{0}^{2}}{l^{2}}\left( \Xi dt-ad\phi \right) ^{2}+%
\frac{r^{2}dr^{2}}{(r^{2}+r_{0}^{2})f(r)}  \nonumber \\
&&+l^{2}f(r)\left( \frac{a}{l^{2}}dt-\Xi d\phi \right) ^{2}+\frac{%
r^{2}+r_{0}^{2}}{l^{2}}dz^{2},  \label{Met3}
\end{eqnarray}%
where $f(r)$ is given in Eqs. (\ref{f2}). The non-vanishing
electromagnetic field components become

\begin{equation}
F_{rt}=-\frac{a}{\Xi l^{2}}F_{r\phi }=\left\{
\begin{array}{cc}
0, & s=0,\frac{1}{2} \\
\frac{-2qa}{\Xi \left( r^{2}+r_{0}^{2}\right) ^{1/2}}, & s=\frac{3}{2} \\
\frac{2(2s-3)qa}{\Xi (2s-1)(r^{2}+r_{0}^{2})^{1/(2s-1)}}, & \text{otherwise}%
\end{array}
\right. .
\end{equation}
The transformation (\ref{Tr}) generates a new metric, because it
is not a permitted global coordinate transformation. This
transformation can be done
locally but not globally. Therefore, the metrics (\ref{Met2}) and (\ref{Met3}%
) can be locally mapped into each other but not globally, and so
they are distinct. Note that this spacetime has no horizon and
curvature singularity. However, it has a conical singularity at
$r=0$. It is notable to mention that for $s=1$, these solutions
reduce to asymptotically AdS magnetic rotating string solutions
presented in \cite{Lem2}.

The mass and angular momentum per unit length of the string when
the boundary $\mathcal{B}$ goes to infinity can be calculated
through the use of Eqs. (\ref{Mastot}) and (\ref{Angtot}). We find
\[
{M}=\frac{\pi }{2}\left[ 3\Xi ^{2}-2\right] m,
\]%
\[
{J}=\frac{3\pi m\Xi l}{2}\sqrt{\Xi ^{2}-1}.
\]%
For $a=0$ ($\Xi =1$), the angular momentum per unit length
vanishes, and therefore $a$ is the rotational parameter of the
spacetime.

Finally, we compute the electric charge of the solutions. To
determine the electric field one should consider the projections
of the electromagnetic
field tensor on special hypersurface. The electric charge per unit length ${Q%
}$ can be found by calculating the flux of the electric field at
infinity, yielding
\begin{equation}
Q=\sqrt{\Xi ^{2}-1}\times \left\{
\begin{array}{cc}
0, & s=0,\frac{1}{2} \\
\frac{3\sqrt{2}lq^{2}}{2\pi \Xi ^{2}}, & s=\frac{3}{2} \\
\frac{-s\left( \frac{8(2s-3)lq}{(2s-1)\Xi }\right)
^{2s-1}}{2^{3s+1}\pi l},
& \text{otherwise}%
\end{array}%
\right. .  \label{chden}
\end{equation}%
It is worth noting that the electric charge is proportional to the
rotation parameter, and is zero for the case of static solutions.
This result is expected since now, besides the magnetic field
along the $\phi $ coordinate, there is also a radial electric
field ($F_{tr}\neq 0$).

\section{Conclusions}

In conclusion, with an appropriate combination of nonlinear
electromagnetic field and Einstein gravity, we constructed a class
of four dimensional magnetic string solutions which produces a
longitudinal magnetic field. These solutions have no curvature
singularity and no horizon, but have conic singularity at $r=0$.
In fact, we showed that near the origin $r=0$, the metric
describes a spacetime which is locally flat but has a conical
singularity at $r=0$ with a deficit angle $\delta \phi=8 \pi \mu$,
where $\mu $ can be interpreted as the mass per unit length of the
string. Also, we investigated the effects of nonlinearity on the
deficit angle and asymptotic behavior of the solutions and found
that for $0<s<\frac{1}{2}$, the solutions are not asymptotically
AdS and for $s<0$\ or $s>\frac{1}{2}$, the asymptotic behavior of
solutions are the same as linear AdS case. In these static
spacetimes, the electric field vanishes and therefore the string
has no net electric charge. Then we added an angular momentum to
the spacetime by performing a rotation boost in the $t-\phi $
plane. For the spinning string, when the rotation parameter is
nonzero, the string has a net electric charge which is
proportional to the magnitude of the rotation parameter. We also
computed the conserved quantities of the solutions by using the
conterterm method.

\acknowledgments{The author would like to thank the anonymous
referee for his enlightening comments. This work has been
supported financially by Research Institute for Astronomy and
Astrophysics of Maragha, Iran.}


\begin{thebibliography}{99}
\bibitem{VachVil91} T. Vachaspati and A. Vilenkin, Phys. Rev. Lett. 67
(1991) 1057.

\bibitem{VilenkinBook} A. Vilenkin and E.P.S. Shellard, \textit{Cosmic
strings and other topological defects}, ( Cambridge University
Press, New York, 1994).

\bibitem{Demi86} M. Born and L. Infeld, Proc. Roy. Soc. Lond. A 144 (1934)
425;

B. Hoffmann, Phys. Rev. D 47 (1935) 877;

M. Demianski, Found. Phys. 16 (1986) 187;

H.P. de Oliveira, Class. Quant. Grav. 11 (1994) 1469;

G.W. Gibbons and D.A. Rasheed, Nucl. Phys. B 454 (1995) 185;

R.G. Cai, D.W. Pang and A. Wang, Phys. Rev. D 70 (2004) 124034;

T.K. Dey, Phys. Lett. B 595 (2004) 484.

\bibitem{DehSed} M.H. Dehghani and H.R. Sedehi, Phys. Rev. D 74 (2006)
124018;

D.L. Wiltshire, Phys. Rev. D 38 (1988) 2445;

M. Aiello, R. Ferraro and G. Giribet, Phys. Rev. D 70 (2004)
104014.

\bibitem{DAH08DH07} M.H. Dehghani, N. Alinejadi, and S.H. Hendi, Phys. Rev.
D 77 (2008) 104025;

M.H. Dehghani and S.H. Hendi, Int. J. Mod. Phys. D 16 (2007) 1829.

\bibitem{AryForVil86} M. Aryal, L.H. Ford and A. Vilenkin, Phys. Rev D 34
(1986) 2263.

\bibitem{Linet87} B. Linet, Phys. Lett. A 124 (1987) 240.

\bibitem{Vil} A. Vilenkin, Phys. Rev. D 23 (1981) 852;\newline
W. A. Hiscock, Phys. Rev. D. 31 (1985) 3288; \newline D. Harari
and P. Sikivie, Phys. Rev. D 37 (1988) 3438;\newline A. D. Cohen
and D. B. Kaplan, Phys. Lett. B 215 (1988) 65; \newline R.
Gregory, Phys. Rev. D. 215 (1988) 663.

\bibitem{Ban} A. Banerjee, N. Banerjee, and A. A. Sen, Phys. Rev. D 53
(1996) 5508; \newline M. H. Dehghani and T. Jalali, Phys. Rev. D
66 (2002) 124014;\newline M. H. Dehghani and A. Khodam-Mohammadi,
Can. J. Phys. 83 (2005) 229.

\bibitem{Bon} W. B. Bonnor, Proc. Roy. S. London A 67 (1954) 225; \newline
A. Melvin, Phys. Lett. 8 (1964) 65.

\bibitem{Lem1} O. J. C. Dias and J. P. S. Lemos, J. High Energy Phys. 01
(2002) 006;

\bibitem{Lem2} O. J. C. Dias and J. P. S. Lemos, Class. Quant. Gravit. 19
(2002) 2265.

\bibitem{Deh2} M. H. Dehghani, Phys. Rev. D 69 (2004) 044024.

\bibitem{Wit2} E. Witten, Nucl. Phys. B 249 (1985) 557; P. Peter, Phys. Rev.
D 49 (1994) 5052.

\bibitem{Moss} I. Moss and S. Poletti, Phys. Lett. B 199 (1987) 34.

\bibitem{Dehhorizonless1} M.H. Dehghani, Phys. Rev. D 64 (2004) 044024.

\bibitem{Dehhorizonless2} M.H. Dehghani, Phys. Rev. D 64 (2004) 064024.

\bibitem{DB07} M.H. Dehghani and N. Bostani, Phys. Rev. D 75 (2007) 084013;

\bibitem{DHWorm} M.H. Dehghani, and S.H. Hendi, Gen. Rel. Grav. (published
online).

\bibitem{VilBanBanSen96} A. Vilenkin, Phys. Rev. D 23 (1981) 852;

A. Banerjee, N. Banerjee and A.A. Sen, Phys. Rev. D 53 (1996)
5508.

\bibitem{LeviCivita19Marder58} T. Levi-Civita, Rend. Reale Accad. Lincei Cl.
Sci. Fis. Mat. Nat. 28 (1919) 3;

L. Marder, Proc. R. Soc. A 244 (1958) 524.

\bibitem{Mukh38Witt85} B.C. Mukherji, Bull. Calcutta Math. Soc. 30 (1938) 95;

E. Witten, Nucl. Phys. B 249 (1985) 557.

\bibitem{DiasLemos02} O.J.C. Dias and J.P.S. Lemos, Class. Quant. Grav. 19
(2002) 2265.

\bibitem{HendiJMP08DSH08} S.H. Hendi, J. Math. Phys. 49 (2008) 082501;

M.H. Dehghani, A. Sheykhi, and S.H. Hendi, Phys. Lett. B 659
(2008) 476;

M.H Dehghani, S.H. Hendi, A. Sheykhi, H. Rastegar Sedehi, JCAP
0702 (2007) 020.

\bibitem{DesSch} S. Deser and A. Schwimmer, Int. J. Mod. Phys. B 8 (1994)
3741;

G. Esposito and C. Stornaiolo, Class. Quant. Grav. 17 (2000) 1989;

G. Esposito and C. Stornaiolo, Nucl. Phys. Proc. Suppl. 88 (2000)
365;

C. Codirla and H. Osborn, Annals Phys. 260 (1997) 91.

\bibitem{EastSing} M. Eastwood and M. Singer, Phys. Lett. A 107 (1985) 73.

\bibitem{Popl08} N.J. Poplawski, Int. J. Mod. Phys. A 23 (2008) 567.

\bibitem{Espo97} G. Esposito, Phys. Rev. D 56 (1997) 2442.

\bibitem{HessMart07MHMar} M. Hassaine and C. Martinez, Phys. Rev. D 75
(2007) 027502;

H. Maeda, M. Hassaine and C. Martinez, [arXiv:gr-qc/08122038].

\bibitem{HendiRast} S.H. Hendi and H.R. Rastegar-Sedehi, Gen. Rel. Grav. 41
(2009) 1355.

\bibitem{BY} J. Brown and J. York, Phys. Rev. D 47 (1993) 1407.

\bibitem{Mal} J. Maldacena, Adv. Theor. Math. Phys. 2 (1998) 231;

E. Witten, Adv. Theor. Math. Phys., 2 (1998) 253;

P. Kraus, F. Larsen and R. Siebelink, Nucl. Phys. B 563 (1999)
259;

O. Aharony, S.S. Gubser, J. Maldacena, H. Ooguri and Y. Oz, Phys.
Rept., 323 (2000) 183;

M.H. Dehghani and R.B. Mann, Phys. Rev. D 64 (2001) 044003;

M.H. Dehghani, Phys. Rev. D 65 (2002) 104030.

\bibitem{Vil2} A. Vilenkin, Phys. Rep. 121 (1985) 263.
\end{thebibliography}
\end{document}